\documentclass[sigconf]{acmart}
\usepackage{threeparttable}
\usepackage{multirow}
\usepackage{makecell}
\usepackage{colortbl}

\usepackage{amsthm,amsmath,amssymb}
\usepackage{subcaption}
\usepackage{graphicx}
\usepackage{enumitem}
\usepackage{bbding}
\usepackage{twemojis}
\usepackage{tablefootnote}
\usepackage{siunitx}
\usepackage[dvipsnames,svgnames,x11names]{xcolor} 
\usepackage{array}
\usepackage{stfloats}
\usepackage[breakable,skins,theorems]{tcolorbox}
\usepackage{tabularx}
\usepackage{makecell}
\usepackage{graphicx}
\usepackage{algorithm}
\usepackage{algpseudocode}
\usepackage{xcolor}       
\usepackage{amsmath}      
\usepackage{amssymb}      
\usepackage{etoolbox}     
\usepackage{listings}
\renewcommand\footnotetextcopyrightpermission[1]{}

\begin{document}


\title{What Would You Click? Personalized Video Thumbnail Generation with Preference-aware Highlight Retrieval}




%

\author{Zhiyu He}
\authornote{These authors contributed equally to this work.\label{first-co}}
\email{hezy18@mails.tsinghua.edu.cn}
\orcid{0000-0003-1291-2739}

\affiliation{%
  \institution{Tsinghua University}
  \city{Beijing}
  \country{China}
}

\author{Zecheng Zhao\textsuperscript{\ref{first-co}}}
\email{uqzzha35@uq.edu.au}
\orcid{0009-0000-3191-0400}
\affiliation{%
  \institution{The University of Queensland}
  \city{Brisbane}
  \country{Australia}
}

\author{Tong Chen}
\orcid{0000-0001-7269-146X}
\email{tong.chen@uq.edu.au}
\affiliation{%
  \institution{The University of Queensland}
  \city{Brisbane}
  \country{Australia}
}

\author{Zi Huang}
\orcid{0000-0002-9738-4949}
\email{huang@itee.uq.edu.au}
\affiliation{%
 \institution{The University of Queensland}
  \city{Brisbane}
  \country{Australia}
}

\author{Yiqun Liu}
\orcid{0000-0002-0140-4512}
\email{yiqunliu@tsinghua.edu.cn}

\affiliation{%
  \institution{Tsinghua University}
  \city{Beijing}
  \country{China}
}

\author{Min Zhang}
\email{z-m@tsinghua.edu.cn}
\authornote{Corresponding Authors}
\orcid{0000-0003-3158-1920}
\affiliation{%
  \institution{Tsinghua University}
  \city{Beijing}
  \country{China}
}

\renewcommand{\shortauthors}{Zhiyu He, et al.}

\begin{abstract}

Video thumbnails are a key factor for attracting user clicks on video platforms, and are increasingly supported by automation. 
However, existing thumbnail generation methods typically produce generic results shared across users, overlooking the diversity of individual preferences. We therefore introduce personalized video thumbnail generation, a novel task that aims to create thumbnails tailored to user-specific preferences. It is challenging in two aspects: (i) identifying visual anchors (i.e., key frames) from each video to guide the generation, which requires a balance between personalization and informativeness that existing highlight detection methods fail to achieve; and (ii) generating personalized thumbnails that are both visually coherent and faithful to the original video. As a response, we propose a two-stage framework that tightly couples preference-aware retrieval with controllable generation. 
In the first stage, a personalized highlight retriever captures fine-grained user-video interactions and incorporates video semantics through summarization, enabling the selection of diverse visual anchors aligned with both user preferences and video contexts. 
In the second stage, a VLM-guided diffusion pipeline transforms these anchors into thumbnails by extracting and injecting semantically grounded visual cues, improving personalization while preserving visual coherence and fidelity.
Experiments on two public datasets show our method delivers state-of-the-art performance compared with both retrieval-based and generative baselines. 
A user study further demonstrates improved click preference, highlighting its effectiveness in enhancing user engagement. The code is available at https://github.com/hezy18/PVTG\label{foot:link}.

\end{abstract}

%


\maketitle

\section{Introduction}
\label{sec:intro}




Video thumbnails serve as the primary visual entry point for video browsing and have a significant impact on user click behavior~\cite{covington2016deep}. 
On modern video platforms, a well-designed thumbnail can dramatically increase a video's click-through rate, making thumbnail selection a critical component of the content delivery pipeline~\cite{liu2015multi,koh2022exploration}. 

\begin{figure}[bt]
    \centering
    \setlength{\abovecaptionskip}{0.2cm}
    
    \includegraphics[width=8.6cm]{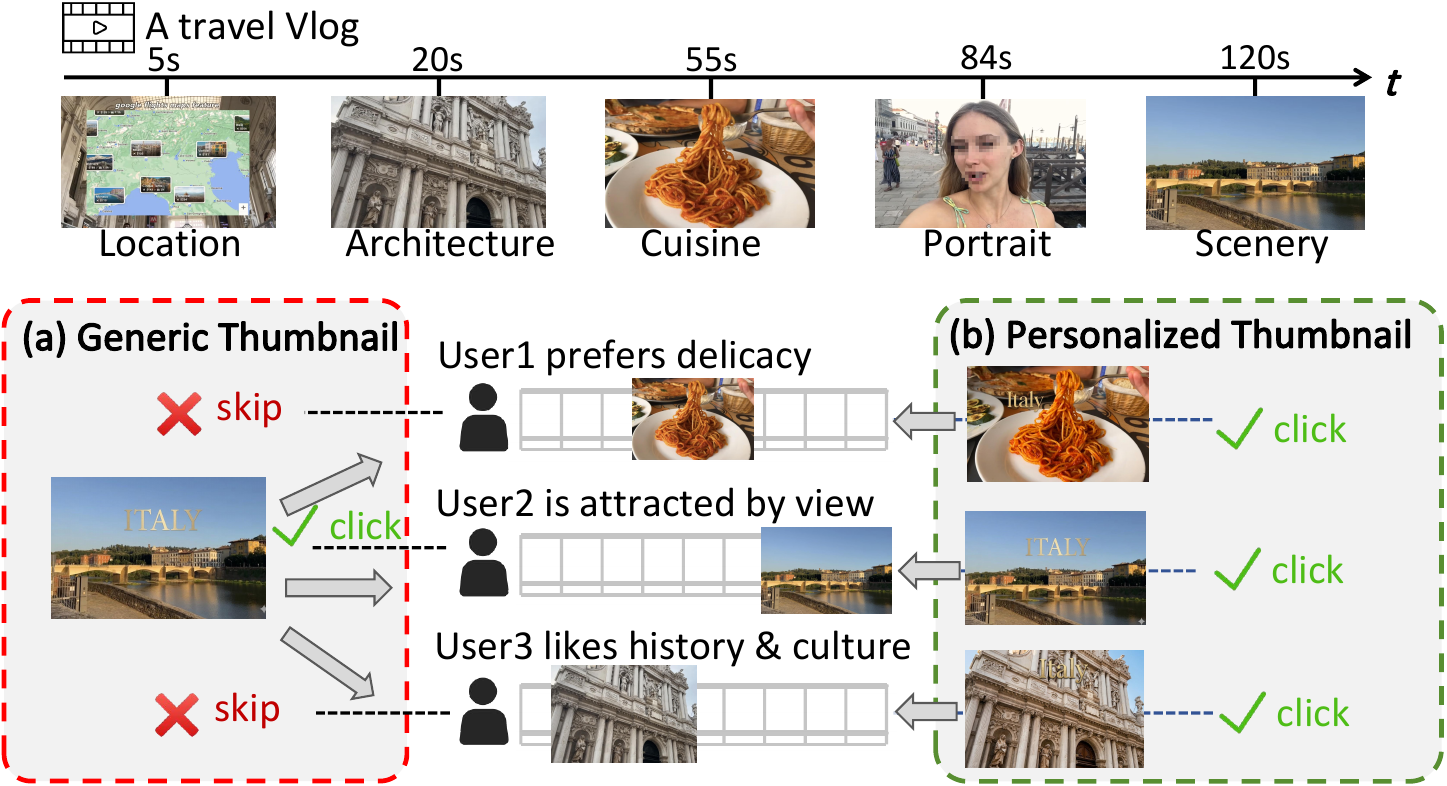}
    \caption{Different users prefer different moments in the same video. A generic thumbnail cannot satisfy all users, while personalized thumbnails highlight user-specific interests to increase click-through rates.}
    \label{fig:motivation} 
    \vspace{-2em}
\end{figure}

To facilitate automation in this process, existing methods focus on selecting thumbnails as key frames from videos that maximize content alignment~\cite{covington2016deep}, visual quality and aesthetics~\cite{song2016click,shimono2020automatic}, or informativeness~\cite{panagiotakis2020personalized}. However, existing methods typically produce a single, generic thumbnail for each video, which is then indiscriminately presented to all users. 
Those homogeneous thumbnails ignore the fact that different users' interests are multifaceted within the same video.
As illustrated in Figure~\ref{fig:motivation}, given a travel video blog (vlog), different users may be attracted to entirely different aspects of it: some may be drawn to the scenic landscapes, while others may find highlights from local food exploration. Clearly, a generic thumbnail cannot simultaneously cater to such diverse preferences, highlighting the need for \textit{personalized} video thumbnails that can fully maximize a video's appeal to individual users. 

With the increasing accessibility and volume of users' interaction records on video platforms, a straightforward fix is resorting to personalized highlight detection~\cite{garcia2018phd,rochan2020adaptive,bhattacharya2022show}, which retrieves the video frame most relevant to a user's historical interests and uses it directly as the thumbnail. 
Nonetheless, a single frame provides only a narrow view of the video, and can hardly guarantee coverage of all the interesting points for a user. In addition, the retrieval-only paradigm makes the visual presentation of the thumbnail entirely dictated by the original footage, with no flexibility to adapt to a target user's preferred style. Ideally, an effective personalized thumbnail should not only encapsulate ``what'' a user likes about the video, but also take into account ``how'' such visual information is presented to the user. Thus, these limitations motivate a paradigm shift from retrieval-only methods to a solution with generation capability. Instead of pinpointing a key frame as the thumbnail for each user, a generative paradigm produces a tailored thumbnail image by binding and rendering the video-specific user interests, offering a larger degree of personalization.


In this paper, we systematically study the pathway towards this paradigm shift, which is subsumed under a new, practical task of \textit{personalized video thumbnail generation}. Unfortunately, in the context of personalized thumbnail generation, two key challenges are currently hindering a workable solution.
Firstly, it is non-trivial to \textbf{effectively identify personalized yet informative visual anchors within each video}. To ensure personalization, locating each user's personal interests within the target video is an indispensable step. This can be generally cast as identifying visual anchors (i.e., highlights/key frames) subject to user preferences \cite{garcia2018phd}, based on which the subsequent generation can be performed. However, when extracting frames from the target video, existing solutions mostly optimize alignment with annotations obtained from historical user interactions \cite{rochan2020adaptive,chen2024personalized}, neglecting their relevance to the video content. Consequently, the identified visual anchors risk overemphasizing a thin slice of the video (e.g., consecutive frames showing user-preferred objectives), leading to substantial loss of contextual information for the downstream generation. 
Secondly, it is necessary to \textbf{maintain control over both visual coherence and faithfulness for the generated thumbnails}. After the personalized visual anchors are identified, they are used to fuel a generative model to produce thumbnails. On the one hand, while image fusion methods~\cite{koik2014image,liu2025infrared} are able to combine key frames via certain layouts, their splicing nature fail to meet the visual coherence demanded by video thumbnails. On the other hand, diffusion-based editing approaches~\cite{ye2023ip,labs2025flux} has more freedom but are prone to hallucinating objects that deviate from the original video, undermining faithfulness. Hence, a mechanism has to be in place to make the generation process remain grounded in the source video, ensuring both visual coherence and content fidelity.

To address the aforementioned challenges, we propose a novel personalized video thumbnail generation framework, which operates in two main stages. 
In our solution, \textbf{Personalized Highlight Retrieval}~(Stage~1) combines user-specific preferences with video context to identify visual anchors that are personalized and informative. A \textit{user-video cross-attention} captures fine-grained correlations between the user and each video shot, and a \textit{semantic summarization} module ensures coverage of the overall content. These signals are input to \textit{Hierarchical Gated Fusion} to assign per-shot interest scores that reflect user preferences while preserving diverse, informative parts of the video, addressing the first challenge. 
To tackle the second challenge of maintaining both visual coherence and faithfulness, \textbf{Personalized Thumbnail Generation}~(Stage~2) generates thumbnails by enhancing the top-ranked frames with salient cues from additional high-interest shots. 
A vision-language model converts these cues into structured editing instructions, which are injected into a \textit{controllable diffusion pipeline} guided by \textit{personalized edit intensity prediction} to produce visually distinct thumbnails tailored to user preferences. A \textit{verify-and-retry} loop ensures fidelity to the source video, preventing hallucinations or incoherent splicing while enabling fine-grained style personalization. As such, for each video, our framework is able to produce meaningfully personalized thumbnails across users.

The contributions of this work are as follows:

\begin{itemize}[nolistsep, leftmargin=*]
    \item We introduce personalized video thumbnail generation, a new task aiming to generate thumbnails tailored to individual user interests. 
    It bridges user preference modeling and controllable image generation, opening up opportunities to optimize user engagement by personalizing how content is presented.
    
    \item 
    We propose a two-stage framework for personalized video thumbnail generation that integrates interest-aware highlight retrieval with diffusion-based editing. 
    The retrieval stage models user preference and video semantics via cross-attention and gated fusion, while the generation stage extracts visual cues from top-ranked frames and performs personalized editing with fidelity judgment. 
    
    \item Extensive experiments on two benchmarks show that our framework leads all competitors in thumbnail quality, content fidelity, and style personalization. A user study further demonstrates the effectiveness of our approach to click preference, indicating its real-world utility. 
      
\end{itemize}
\section{Related Work}
\label{sec:related}

\subsection{Video Thumbnail Extraction} 

Automated video thumbnail extraction mainly treats the task as key-frame selection.
\citet{covington2016deep} train deep neural networks on click-through data to rank frames by engagement, linking frame quality to viewer attraction. Aesthetic-driven methods~\cite{song2016click,shimono2020automatic} score frames covering sharpness, composition balance, and face saliency. 
Summarization-based approaches~\cite{panagiotakis2020personalized} reframe thumbnail extraction as video summarization, choosing the frame that maximizes coverage of key events. 
Beyond single-frame selection, some work assembles multiple visual elements into a composite thumbnail layout, stitching together representative frames and foreground objects to define layouts~\cite{hsu2025postero}. \citet{yang2023toward} incorporate user perception into this design, showing that perceptually salient composites increase engagement. 
However, these methods produce identical thumbnails for all viewers, failing to capture users’ diverse content interests or adapt to their preferred visual style.

\begin{figure*}[htbp]
    \centering
    \setlength{\abovecaptionskip}{0.1cm}
    \includegraphics[width=18.0cm]{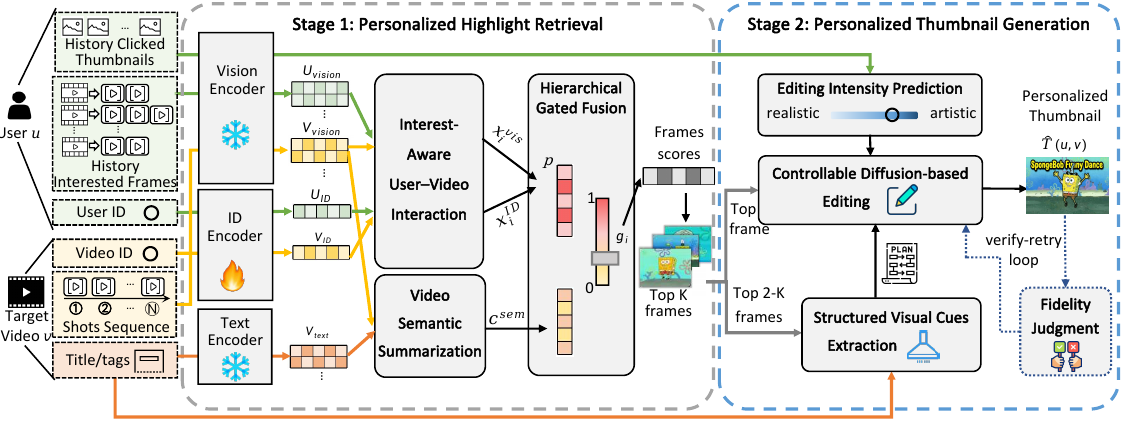}
    \caption{Overview framework with two stages: personalized highlight retrieval and thumbnail generation.}
    \label{fig:overview} 
    \vspace{-1em}
\end{figure*}

\subsection{Personalized Highlight Detection} 

Personalized video highlight detection relates to automated thumbnail generation, as both aim to select frames matching user preferences~\cite{rochan2020adaptive,wei2022learning}, and our Stage~1 builds on this task to identify frames most aligned with a user’s historical interests. 
PHD-GIFs~\cite{garcia2018phd} is the first personalized video highlight detection technique that extracts highlight predictions guided by subjects' manually created GIFs. 
Models are optimized based on individuals' preferred clips to guide a content-based highlight detection network~\cite{rochan2020adaptive,chen2021pr}.  
ShowMe~\cite{bhattacharya2022show} builds upon this approach by employing a multi-head attention mechanism, achieving improved performance. 
More recently, KuaiHL~\cite{deng2024multimodal} propose a multimodal Transformer for live streaming highlight prediction that jointly models visual, audio, and textual streams, demonstrating strong performance in industrial short-video settings.
Despite their progress, existing methods mainly infer highlights by matching candidate segments with users’ past annotations, focusing primarily on user preference modeling. 

\subsection{Image Editing and Generation}





Image generation has been extensively studied in layout and poster design.
LayoutLLM~\cite{luo2024layoutllm} and LayoutDiffuse~\cite{cheng2023layoutdiffuse} generate structured layouts under semantic constraints, while PosterLLaVa~\cite{yang2024posterllava} and PosterO~\cite{hsu2025postero} optimize poster arrangements through multimodal modeling and iterative refinement. These methods operate over fixed design primitives for general visual tasks, without conditioning on individual user preferences.
Another line of work focuses on layout-aware image composition and diffusion-based editing. NoiseCollage~\cite{shirakawa2024noisecollage} and LAMIC~\cite{chen2026lamic} produce visually coherent composites under spatial constraints, while IP-Adapter~\cite{ye2023ip} and FLUX.1~\cite{labs2025flux} enable reference-guided or fine-grained controllable editing. 
However, layout-based methods suffer from visual incoherence, while unconstrained diffusion editors may hallucinate content that deviates from the source video, undermining faithfulness. 

\section{Method}
\label{sec:method}

\subsection{Problem Definition}

Let $\mathcal{U}$ denote the set of users and $\mathcal{V}$ the set of videos. 
Each video $v \in \mathcal{V}$ is represented as a sequence ${s_1, \ldots, s_N}$, where each $s_i$ denotes the representative frame of an automatically detected shot (see Appendix for details).  
A user $u \in \mathcal{U}$ has two types of historical signals: 
(1) clicked thumbnails $\mathcal{H}^T_u$, and 
(2) sequence-level interaction history $\mathcal{H}^S_u = \{(v_1, S_1^+), (v_2, S_2^+), \ldots\}$, where $S_k^+ \subseteq \{s_1^{(k)}, \ldots, s_{N_k}^{(k)}\}$ denotes frames aligned with user interests.  
We adopt curated highlight annotations~\cite{garcia2018phd} as supervision, which naturally generalizes to industrial platforms via implicit engagement signals such as repeated viewing~\cite{zhan2022deconfounding, he2025segmm}, commenting~\cite{gu2024holistic,yang2016video}, or content sharing~\cite{yang2016time}. 

Given a user $u$ and a new video $v$ unseen by $u$, the goal is to generate a personalized thumbnail $\hat{T}(u, v)$ reflecting $u$'s interest while remaining faithful to $v$. 
The generated thumbnail is expected to jointly reflect personalized retrieved content and personalized generation style, thereby better aligning with user interests and improving click-through probability. 


\subsection{Model Overview}

Figure~\ref{fig:overview} illustrates the proposed method that consists of two stages: personalized highlight retrieval and personalized thumbnail generation. 
It takes as input the frame sequence $\{s_1, \ldots, s_N\}$, user history $(\mathcal{H}^T_u, \mathcal{H}^S_u)$, and video metadata~(e.g., titles or tags), and produces a personalized thumbnail. 



\noindent \textbf{Stage~1: Personalized Highlight Retrieval.}
We estimate the interest score $\hat{y}_i = \phi(s_i \mid u, v; \Theta)$ to frame $i$ by jointly modeling user preference and video semantics. 
User and video embeddings of different modalities are jointly refined through cross-attention to extract fine-grained preference signals, while visual features are combined with textual context to encode global semantics. 
A hierarchical fusion module balances user preference with content coverage, producing scores and retrieving top-ranked frames.

\noindent \textbf{Stage~2: Personalized Thumbnail Generation.}
Given retrieved frames, we generate a coherent thumbnail by injecting structured visual cues into a base frame. 
A vision-language model produces structured editing instructions, which are executed via controllable diffusion-based editing. 
Editing strength is modulated by personalized intensity prediction derived from $\mathcal{H}^T_u$. The verify-and-retry loop ensures coherence and adherence to the source video. 
This design naturally couples personalized retrieval with faithful thumbnail generation, resulting in output thumbnail $\hat{T}(u, v)$ that reflects user interests while maintaining visual consistency.




\subsection{Stage~1: Personalized Highlight Retrieval}
\label{sec:retrieval}


Personalized Highlight Retrieval relies on capturing fine-grained user preferences and the video's semantic structure. It estimates per-frame interest score through Interest-Aware User-Video Interaction, Video Semantic Summarization, and Hierarchical Gated Fusion. 

\subsubsection{Interest-Aware User-Video Interaction}

To capture fine-grained personalized interests, we introduce the Interest-Aware User–Video Interaction module, which distinguishes user history from video frame representations: user history encodes accumulated preferences, while video frames capture momentary visual content. Treating them uniformly can dilute critical user–video interactions essential for highlight prediction. Our design builds on cross-attention mechanisms~\cite{he2025segmm} to maintain separate user and video representations and update them jointly.

Each video has $N$ frames $s_i$ represented in two modalities: visual embeddings $\mathbf{V}_{\text{vision}} \in \mathbb{R}^{N \times d}$ and ID embeddings $\mathbf{V}_{\text{ID}} \in \mathbb{R}^{N \times d}$; user history embeddings are $\mathbf{U}_{\text{vision}} \in \mathbb{R}^{M \times d}$ and $\mathbf{U}_{\text{ID}} \in \mathbb{R}^{1 \times d}$. Cross-attention updates:
\begin{align}
\mathbf{A}_{VU}^{(l)} &= \mathrm{FFN}(\mathbf{V}^{(l)}) \cdot \mathrm{FFN}(\mathbf{U}^{(l)})^\top / \sqrt{d}, \\
\mathbf{A}_{VV}^{(l)} &= \mathrm{FFN}(\mathbf{V}^{(l)}) \cdot \mathrm{FFN}(\mathbf{V}^{(l)})^\top / \sqrt{d}, \\
\mathbf{V'}^{(l)} &= \mathrm{Softmax}({A}_{VU}^{(l)} \oplus {A}_{VV}^{(l)}) \cdot \left(\mathit{FFN}(V^{(l)}) \oplus FFN(U^{(l)})\right).
\end{align}
A symmetric update procedure is applied to $\mathbf{U}^{(l)}$. Residual connections and layer normalization ensure each frame integrates user history and intra-video context. The resulting embeddings $\mathbf{x}_i^{\text{vis}}$ and $\mathbf{x}_i^{\text{ID}}$ are passed to the decoder for scoring.

\subsubsection{Video Semantic Summarization}

Visual reasoning alone cannot distinguish videos with similar frames but different topics. To emphasize frames central to the video’s topic, we incorporate global context from titles and tags. 
Let the Text Encoder produce $\mathbf{V}_{\text{text}} \in \mathbb{R}^{d}$ from concatenated title and tag tokens. We compute a query $\mathbf{q}_t = \mathbf{W}_p \mathbf{t}_v$ and attend over visual frames:
\begin{align}
\mathbf{c}^{\text{sem}} = \mathrm{FFN}\Big(\mathrm{LayerNorm}(\mathbf{q}_t + \mathrm{CrossAttn}(\mathbf{q}_t, \mathbf{V}_{\text{vision}}^{(L)}))\Big) \in \mathbb{R}^d,
\end{align}
producing a frame-aligned semantic context vector that grounds the title and tags to frames.



\subsubsection{Hierarchical Gated Fusion}

Each frame has three signals: 
$\mathbf{x}_i^{\text{vis}}$ (visual-personalized), $\mathbf{x}_i^{\text{ID}}$ (collaborative ID), and $\mathbf{c}^{\text{sem}}$ (semantic context). Different frames rely differently on user preference and semantic relevance. The Hierarchical Gated Fusion module integrates these signals in two steps: first, it combines the user-specific features, then it adaptively balances user preference against semantic context per dimension.

We first merge the visual and ID modalities into a compact user interest vector that summarizes complementary appearance-based and collaborative signals, representing “what the user likes” for frames $s_i$:
\begin{equation}
\mathbf{p}_i = \mathrm{LayerNorm}\big(\mathbf{W}_p [\mathbf{x}_i^{\text{vis}}; \mathbf{x}_i^{\text{ID}} + \mathbf{b}_p\big) \in \mathbb{R}^d.
\end{equation}
An adaptive gate then modulates the contribution of user preference and semantic context, and the two are fused via a gate-modulated convex combination:
\begin{equation}
\mathbf{g}_i = \sigma(\mathbf{W}_g [\mathbf{p}_i; \mathbf{c}^{\text{sem}}] + \mathbf{b}_g), \quad
\mathbf{r}_i = \mathrm{LayerNorm}(\mathbf{g}_i \odot \mathbf{p}_i + (1-\mathbf{g}_i) \odot \mathbf{c}^{\text{sem}}),
\end{equation}
where $\mathbf{g}_i$ adaptively controls the influence of user preference versus semantic coverage. Finally, the per-frame interest score is computed as $\hat{y}_i = \sigma(\mathrm{MLP}(\mathbf{r}_i))$.

\subsubsection{Training Process}

In each video, only a small set of frames are highlights ($y_i = 1$), while most are non-highlights ($y_i = 0$). To prevent the model from being dominated by abundant negative frames, we adopt a focal loss that emphasizes the rare but semantically critical positive frames. For each frame $s_i$ with predicted score $\hat{y}_i$ and label $y_i$, the per-frame loss is: 
\begin{equation}
  \mathcal{L}_{\text{focal}}^{i}
  = \alpha_i \,(1 - p_{t,i})^{\gamma}\,\mathcal{L}_{\text{BCE}}(\hat{y}_i,\, y_i),
  \label{eq:focal}
\end{equation}
where $p_{t,i} = \hat{y}_i$ if $y_i = 1$ and $1 - \hat{y}_i$ otherwise with class-balancing weight $\alpha_i = \alpha\, y_i + (1-\alpha)(1-y_i)$. 

The overall loss averages $\mathcal{L}_{\text{focal}}^{i}$ over all frames in the training set. During inference, the model computes $\hat{y}_i$ for each frame and selects the top-ranked frame as the candidate for thumbnail generation.

\subsection{Stage~2: Personalized Thumbnail Generation} 
\label{sec:generation}

Our personalized highlight retrieval model identifies the video frame most aligned with each user's interests. However, the top-ranked key frame alone is insufficient as a thumbnail. It is a single uncontrolled snapshot that offers a narrow view of the video, and its visual composition is entirely determined by the original footage. Enhancing it through generation is necessary, yet this introduces a core difficulty: the generated thumbnail must be personalized and remain visually coherent and faithful to the source video. 

This core problem can be framed from three perspectives. (i) \textbf{What to edit}: the injected visual cues should be semantically meaningful and derived from the user's preferred moments, not arbitrarily chosen. 
Naive layout-based methods~\cite{hsu2025postero,yang2023toward} stitch visual regions rigidly without semantic understanding and cannot condition on user preference. 
(ii) \textbf{How much to edit}: the degree of visual modification should reflect individual taste, as some users prefer near-original frames while others favor heavily stylized thumbnails. Yet current diffusion editing tools~\cite{chen2026lamic,labs2025flux} apply a uniform editing intensity with no per-user calibration. 
(iii) \textbf{How to ensure fidelity}: the generated thumbnail must stay grounded in the source video to preserve visual coherence and content accuracy. Without structured guidance, however, diffusion models can hallucinate details that deviate from the original footage, undermining faithfulness. 

To address these challenges, we develop a personalized pipeline that extracts visual cues from user-preferred frames, predicts per-user editing intensity, and applies controllable diffusion-based editing with closed-loop judgment. 

\vspace{-1em}
\subsubsection{Structured Visual Cue Extraction} 
\label{structured editing plan}
The first challenge is 
deciding what to edit: the injected cues must be semantically meaningful and traceable to the user's preferred moments. The top-ranked frames from our retrieval model implicitly encode user visual preferences. 
To translate these implicit preferences into explicit editing instructions for the downstream diffusion model, we extract structured visual cues that specify what to edit and where.

Specifically, the cleaned key frames (with overlaid text removed), together with the video's title and tags, are passed to a vision-language model (Qwen3.5-27B~\cite{qwen_qwen3_5_27b_hf}). The model takes the ranked frames and video metadata as input, with the objective of producing a structured editing plan for the top-ranked base frame. It outputs three components: (i) a content summary of the base frame describing its dominant subject and scene; (ii) a fusion decision that assesses semantic similarity across ranked frames and determines whether complementary cues from supporting frames should be integrated into the base frame; and (iii) a list of concrete editing operations covering topic visualization, composition adjustment, subject emphasis, and background control. The detailed prompt design is provided in Appendix. This structured plan ensures that all downstream editing operations are grounded in user-preferred visual content rather than arbitrary design choices.

\vspace{-1em}
\subsubsection{Editing Intensity Prediction} \label{Intensity Prediction}
Beyond deciding what to edit, the pipeline 
predicts how much to edit, since users differ in their tolerance for visual modification. To introduce personalization into the editing process
, we develop a training-free intensity predictor with offline calibration and online prediction.

We first define four style axes that characterize how an edited thumbnail deviates from its source frame: \textit{subject pop-out} (strength of the focal point), \textit{subject dominance} (how tightly the subject is framed), \textit{semantic cue density} (number of supporting elements that explain the topic), and \textit{scene stylization} (mood lighting, color contrast, and background design). Each axis $j$ is described by a pair of 
contrastive natural-language prompts $(t_j^{\text{low}}, t_j^{\text{high}})$ representing its low and high extremes.

We then define four editing intensity levels that follow a strict cumulative hierarchy, progressing from realistic preservation to artistic transformation. \textit{Low} permits only surface treatments such as subject emphasis and background blur. \textit{Medium} additionally allows composition adjustments. \textit{High} further enables fusion of elements from supporting frames. \textit{Extra-high} adds scene transformation including background replacement and topic-level visual elements. Each level filters the structured editing plan from Section~\ref{structured editing plan} to retain only the operations it permits.



\noindent \textbf{Offline calibration.} For each intensity level $l \in \mathcal{L}$, we generate a set of reference thumbnails $\{x_k^{(l)}\}_{k=1}^{K_l}$ from curated source frames using the diffusion-based editing model. Each reference is encoded by CLIP and scored against the axis prompts to yield a 4D style vector: $\psi_j(x) = \mathrm{sim}(\mathbf{z}_x,\, \mathbf{t}_j^{\mathrm{high}}) - \mathrm{sim}(\mathbf{z}_x,\, \mathbf{t}_j^{\mathrm{low}})$ for $j = 1,\dots,4$, where $\mathbf{z}_x$ is the CLIP image embedding of thumbnail $x$. The style prototype for level $l$ is the centroid: $\mathbf{p}^{(l)} = \frac{1}{K_l}\sum_{k=1}^{K_l} \mathbf{\psi}(x_k^{(l)})$.

\noindent \textbf{Online prediction.} 
Given a user $u$ with historical clicked thumbnails $\mathcal{H}_u^T = \{h_1, \dots, h_M\}$, each thumbnail is scored by the same style scoring function to produce a user-level style profile: $\mathbf{q}^{(u)} = \frac{1}{M}\sum_{m=1}^{M} \mathbf{\psi}(h_m)$, where $\mathbf{\psi}(\cdot)$ denotes the 4D style vector. The predicted intensity is determined by nearest-neighbor matching: $\lambda_u = \arg\min_{l \in \mathcal{L}} \|\mathbf{q}^{(u)} - \mathbf{p}^{(l)}\|_2$, which filters the structured editing plan accordingly 
This ensures that the visual transformation is personalized to the user's style preferences.

\subsubsection{verify-and-retry loop.}
\label{verify-retry-loop}
 Even with personalized content and user-calibrated editing intensity, diffusion models can still generate artifacts, such as unnatural compositions or hallucinated objects that deviate from the source video. 
 

 To address this, we develop a VLM-as-judge mechanism that serves as the backbone of a verify-and-retry loop. Specifically, the VLM judge takes the intensity-aware editing plan from Section \ref{Intensity Prediction} and the generated thumbnail as input, and outputs a fidelity score assessing whether the thumbnail faithfully follows the editing instructions. If the score falls below a threshold, the pipeline re-invokes the diffusion editor with resampled noise. This loop continues for up to $N$ attempts, after which the highest-scoring candidate is selected. This closed-loop design ensures that the personalized editing instructions are faithfully executed.
\section{Offline Quantitative Experiments}
\label{sec:offline}

To validate our contributions and demonstrate the effectiveness of the proposed framework, we conduct extensive offline experiments. (i) We evaluate both highlight retrieval accuracy from Stage 1, and whether the produced thumbnails are aesthetically appealing, faithful to the source video, and personalized across users from Stage 2. (ii) We isolate the contribution of each key component to quantify their individual impact. (iii) We examine how key hyperparameters of our framework affect generation quality and personalization.

\subsection{Experimental Setup}


\subsubsection{Dataset}

We evaluate our framework on two public datasets with user highlight annotations. \textbf{EEGSVRec}~\cite{zhang2024eeg} derives per-user highlight labels from EEG-based engagement signals, and \textbf{PHD$^2$}~\cite{garcia2018phd}, where over 10 thousand users create GIF timestamps from YouTube videos, serving as explicit highlight annotations. Both datasets use an 8:1:1 user-based split. We use the full test set for Retrieval experiments. 
For all thumbnail generation experiments, we use a sampled subset with 30 users $\times$ 10 videos for EEGSVRec, and 800 users for \textbf{PHD$^2$} due to per-instance inference cost.

\subsubsection{Baselines}
\noindent \textbf{(i) Key Frame Selection Methods.} These methods select a single frame as the thumbnail without generation. We include five personalized highlight detector: \textbf{HighlightMe}~\cite{bhattacharya2021highlightme}, \textbf{PAC-Net}~\cite{garcia2018phd}, \textbf{PR-Net}~\ \cite{chen2021pr}, \textbf{ShowMe}~\cite{bhattacharya2022show}, and \textbf{KuaiHL}~\cite{deng2024multimodal}, whose top-1 retrieved frame serves as the thumbnail. We additionally include \textbf{ClickorNot}~\cite{song2016click}, a non-personalized thumbnail selector that produces a single generic thumbnail shared across all users. 

\noindent \textbf{(ii) Layout Generation Methods.} These methods compose multiple visual elements into structured thumbnails. We include \textbf{HPCVTG}~\cite{yang2023toward} and \textbf{PosterO}~\cite{hsu2025postero}, both of which require additional thumbnail-specific information and visual inputs. For fair comparison, we supply them with frames retrieved by our Stage 1 model. Detailed descriptions of methods are provided in the Appendix.

\subsubsection{Evaluation Metrics}
\noindent \textbf{(i) Highlight Retrieval Metrics.} We adopt \textbf{mAP} for overall ranking quality, \textbf{Recall@$k$} ($k \in \{1, 3, 5, 10\}$) for top-$k$ hit rate, and \textbf{nMSD} for ranking displacement. 
\noindent \textbf{(ii) Thumbnail Generation Metrics.} We evaluate generated thumbnails along three dimensions.
\textit{\textbf{Thumbnail Quality}} measures visual appeal using \textbf{NIMA}~\cite{talebi2018nima} and \textbf{CLIP-IQA}~\cite{CLIP-IQA}.
\textit{\textbf{Content Fidelity}} measures whether the thumbnail faithfully represents the user's highlighted content using \textbf{LPIPS}~\cite{LPIPS} and \textbf{DINO-Sim}~\cite{DINO}. For each sample, we compute average similarity between the key frame selected by each method and all ground-truth highlight key frames annotated for that user and video.
\textit{\textbf{Style Personalization}} measures whether different users receive visually distinct thumbnails for the same video. We propose \textbf{IUD} (Inter-User Dissimilarity), computing the average pairwise CLIP embedding dissimilarity across users' thumbnails for a shared video.
IUD requires multiple users to receive thumbnails for the same video. In PHD$^2$, nearly every user is recommended a distinct video set, yielding very few shared user--video pairs. This sparsity makes pairwise dissimilarity unreliable, so we report IUD only on EEGSVRec. 
Detailed definitions are in Appendix.

\subsubsection{Implementation Details}
\label{implementation}
For Stage~1, we implement our model in PyTorch~\cite{imambi2021pytorch}. Visual features are extracted using CLIP~\cite{radford2021learning}. The model uses an embedding dimension of $d=256$, $L=2$ cross-attention layers. The maximum user history length is 20. We train with the Adam optimizer~\cite{kingma2014adam} at a learning rate of $1\times10^{-4}$ and apply early stopping with a patience of 10 epochs based on mAP on the validation set.
For Stage~2, we use Qwen3.5-27B~\cite{qwen_qwen3_5_27b_hf} as the vision-language model for structured editing planning and fidelity judgment, and Qwen-Image-Edit-2511~\cite{wu2025qwenimage} as the diffusion-based editor. The verify-and-retry loop runs for up to N=1 attempts. Style axis prototypes are calibrated from $K=50$ reference thumbnails per intensity level. All experiments are conducted on one NVIDIA H100 GPU, with the thumbnail generation taking around 210 seconds per video. For all offline thumbnail generation metrics, we evaluate the edited thumbnails without overlaid text captions, since large text elements would dominate perceptual quality measurements, obscuring the actual visual fidelity of the thumbnail.



\subsection{Overall Performance}

\subsubsection{Personalized Highlight Shot Retrieval Performance}
\label{sec:stage1results}

\newcolumntype{Y}{>{\centering\arraybackslash}X}
\begin{table*}[htbp]
\definecolor{lightpurple}{RGB}{248, 231, 242}
\centering
\setlength{\abovecaptionskip}{0.1cm}
\caption{Performance on personalized highlight frame retrieval. The top and second-best results are highlighted in bold and underlined, respectively. Higher is better for mAP and Recall, while lower is better for NMSD.}
\label{tab:main_results}
\setlength{\aboverulesep}{0pt}
\setlength{\belowrulesep}{0pt}
\setlength{\tabcolsep}{3.5pt} 
\begin{tabularx}{\textwidth}{l YYYYYY | YYYYYY}
\toprule
\multicolumn{1}{c}{\multirow{2}{*}{\textbf{Model}}} & \multicolumn{6}{c|}{\textbf{EEGSVRec}} & \multicolumn{6}{c}{\textbf{PHD$^2$}} \\
\cmidrule(lr){2-7} \cmidrule(lr){8-13}
& mAP $\uparrow$ & R@1 $\uparrow$ & R@3 $\uparrow$ & R@5 $\uparrow$ & R@10 $\uparrow$ & \multicolumn{1}{c|}{NMSD $\downarrow$} & mAP $\uparrow$ & R@1 $\uparrow$ & R@3 $\uparrow$ & R@5 $\uparrow$ & R@10 $\uparrow$ & \multicolumn{1}{c}{NMSD $\downarrow$} \\
\midrule
HighlightMe~\cite{bhattacharya2021highlightme}  & 0.21 & 0.13 & 0.32 & 0.42 & 0.61 & 52.41\% & 0.11 & 0.08 & 0.17 & 0.24 & 0.41 & 64.21\% \\ 
PR-Net~\cite{chen2021pr}       & 0.30 & 0.20 & 0.52 & 0.65 & 0.80 & 51.52\% & 0.15 & 0.10 & 0.21 & 0.31 & 0.51 & 58.63\% \\
PAC-Net~\cite{garcia2018phd}      & 0.38 & 0.30 & 0.66 & 0.75 & 0.86 & 34.66\% & 0.18 & 0.12 & 0.26 & 0.37 & 0.56 & 55.20\% \\
ShowMe~\cite{bhattacharya2022show} & \underline{0.46} & \underline{0.20} & \underline{0.52} & \underline{0.70} & \underline{0.89} & \underline{37.56\%} & \underline{0.23} & \underline{0.16} & \underline{0.34} & \underline{0.47} & \underline{0.68} & \underline{45.21\%} \\
KuaiHL~\cite{deng2024multimodal}       & 0.24 & 0.20 & 0.43 & 0.58 & 0.74 & 46.25\% & 0.22 & 0.15 & 0.33 & 0.44 & 0.65 & 47.50\% \\
\midrule
\rowcolor{lightpurple}
\textbf{Ours} & \textbf{0.60} & \textbf{0.39} & \textbf{0.67} & \textbf{0.78} & \textbf{0.92} & \textbf{31.99\%} & \textbf{0.28} & \textbf{0.19} & \textbf{0.39} & \textbf{0.51} & \textbf{0.71} & \textbf{38.42\%} \\
\bottomrule
\end{tabularx}
\vspace{-1em}
\label{tab:stage1_results}
\end{table*}

\begin{table*}[t]
  \definecolor{lightpurple}{RGB}{248, 231, 242}
  \centering
\setlength{\abovecaptionskip}{0.1cm}
  \caption{Performance comparison of thumbnail generation across quality, fidelity, and style personalization. The top and second-best results are highlighted in bold and underlined, respectively. Higher is better for NIMA, CLIP-IQA and IUD, while lower is better for LPIPS.}
  \label{tab:model_comparison}
  \setlength{\aboverulesep}{0pt}
  \setlength{\belowrulesep}{0pt}
  \setlength{\tabcolsep}{1.6pt} 
  \begin{tabular}{cc ccccc | cccc}
    \toprule
    \multicolumn{1}{c}{\multirow{3}{*}{\textbf{Category}}} & \multicolumn{1}{c}{\multirow{3}{*}{\textbf{Model}}} & \multicolumn{5}{c|}{\textbf{EEGSVRec}} & \multicolumn{4}{c}{\textbf{PHD$^2$}} \\
    \cmidrule(lr){3-7} \cmidrule(lr){8-11}
    & & \multicolumn{2}{c}{Thumbnail Quality} & \multicolumn{2}{c}{Content Fidelity} & \multicolumn{1}{c|}{Style Personalization} & \multicolumn{2}{c}{Thumbnail Quality} & \multicolumn{2}{c}{Content Fidelity} \\
    \cmidrule(lr){3-4} \cmidrule(lr){5-6} \cmidrule(lr){7-7} \cmidrule(lr){8-9} \cmidrule(lr){10-11}
    & & NIMA $\uparrow$ & CLIP-IQA $\uparrow$ & LPIPS $\downarrow$ & DINO-Sim $\uparrow$ & IUD $\uparrow$ & NIMA $\uparrow$ & CLIP-IQA $\uparrow$ & LPIPS $\downarrow$ & DINO-Sim $\uparrow$ \\
    \midrule
    \multirow{3}{*}{
      \begin{tabular}[c]{@{}c@{}}\textit{Key Frame} \\ \textit{Selection}
    \end{tabular}} & ClickorNot~\cite{song2016click} & 4.86 & 0.72 & 0.57 & \underline{0.49} & 0.00 & 4.35 & 0.67 & 0.61 & 0.45 \\
    & KuaiHL~\cite{deng2024multimodal} & 4.77 & 0.71 & \underline{0.56} & 0.47 & \underline{0.03} & 4.26 & \underline{0.70} & \textbf{0.58} & \textbf{0.48} \\
    & ShowMe~\cite{bhattacharya2022show} & 4.82 & \underline{0.73} & \textbf{0.45} & \textbf{0.56} & 0.00 & 4.25 & 0.69 & \underline{0.59} & 0.45 \\
    \midrule
    \multirow{2}{*}{
      \begin{tabular}[c]{@{}c@{}}\textit{Layout} \\ \textit{Generation}
    \end{tabular}} & PosterO~\cite{hsu2025postero} & 4.92 & 0.70 & \textbf{0.45} & \textbf{0.56} & 0.00 & 4.58 & 0.68 & \textbf{0.58} & \underline{0.47} \\
    & hpcvtg~\cite{yang2023toward} & \underline{5.01} & 0.71 & \textbf{0.45} & \textbf{0.56} & 0.00 & \underline{4.77} & 0.61 & \textbf{0.58} & \underline{0.47} \\
    \midrule
    \rowcolor{lightpurple}
    \begin{tabular}[c]{@{}c@{}}\textit{Personalized} \\ \textit{Generation}
    \end{tabular} & \textbf{Ours} & \textbf{5.19} & \textbf{0.75} & \textbf{0.45} & \textbf{0.56} & \textbf{0.15} & \textbf{5.11} & \textbf{0.74} & \textbf{0.58} & \underline{0.47} \\
    \bottomrule
  \end{tabular}
\label{tab:stage2_results}
\end{table*}

\begin{table}[htbp]
\definecolor{lightyellow}{RGB}{255, 249, 192}
\definecolor{lightpurple}{RGB}{248, 231, 242}
\definecolor{deltared}{RGB}{255,150,150}
\definecolor{deltagreen}{RGB}{0, 150, 0}
\newcommand{\res}[2]{#1 \footnotesize{\textcolor{deltared}{(#2)}}}
\newcommand{\respos}[2]{#1 \footnotesize{\textcolor{deltagreen}{(+#2)}}}
\newcommand{\resneutral}[1]{#1 \footnotesize{\textcolor{gray}{(0.00)}}}
\setlength{\abovecaptionskip}{0.1cm}
\caption{Ablation study of key components of our model. \textbf{PHR} denotes Personalized Highlight Retrieval, \textbf{PTG} denotes Personalized Thumbnail Generation, and \textbf{PIP} denotes Personalized Intensity Prediction.}
\label{tab:ablation_eegsvrec}
\centering
\setlength{\aboverulesep}{0pt}
\setlength{\belowrulesep}{0pt}
\small 
\setlength{\tabcolsep}{1pt}
\begin{tabular}{l|ccccc}
\toprule
\multicolumn{1}{l|}{\multirow{3}{*}{\textbf{Model}}} & \multicolumn{5}{c}{\textbf{EEGSVRec}} \\
\cmidrule(lr){2-6}
 & \multicolumn{2}{c}{\makecell{Thumbnail \\ Quality}} & \multicolumn{2}{c}{\makecell{Content \\ Fidelity}} & \makecell{Style \\ Personalization} \\
\cmidrule(lr){2-3} \cmidrule(lr){4-5} \cmidrule(lr){6-6}
 & NIMA $\uparrow$ & CLIP-IQA $\uparrow$ & LPIPS $\downarrow$ & DINO-Sim $\uparrow$ & IUD $\uparrow$ \\
\midrule
\rowcolor{lightpurple}
\textbf{Ours} & \textbf{5.19} & \textbf{0.75} & \textbf{0.45} & \textbf{0.56} & \textbf{0.15} \\
~~w/o PHR & \res{4.95}{-0.24} & \res{0.71}{-0.04} & \res{0.57}{+0.12} & \res{0.48}{-0.08} & \res{0.10}{-0.05} \\
~~w/o PTG & \res{4.82}{-0.37} & \res{0.71}{-0.04} & \resneutral{0.45} & \resneutral{0.56} & \res{0.00}{-0.15} \\
~~w/o PIP & \res{4.88}{-0.31} & \res{0.70}{-0.05} & \resneutral{0.45} & \resneutral{0.56} & \res{0.10}{-0.05} \\
\bottomrule
\end{tabular}
\vspace{-10pt}
\end{table}

Table~\ref{tab:stage1_results} presents personalized highlight retrieval results on EEGSVRec and PHD$^2$. Our method outperforms all baselines, showing that modeling user history and joint user-video interactions is crucial for accurate shot ranking. By combining collaborative filtering with visual content, our approach captures both fine-grained user interests and global semantics, retrieving accurate personalized highlights and providing editable, user-aligned content for Stage~2.

\subsubsection{Thumbnail Generation Performance}
\label{sec:stage2results}
Table~\ref{tab:stage2_results} compares thumbnail quality, content fidelity, and style personalization across all methods. We select the top-2 retrieval baselines from Table~\ref{tab:stage1_results}, KuaiHL and ShowMe, and use their top-1 retrieved frames as personalized thumbnails. For thumbnail quality, key frame selection methods score lowest because raw video frames often contain visual clutter such as overlaid text and suboptimal composition. Layout methods partially improve quality through structured composition but remain limited by rigid template-based stitching. Our structured visual cue extraction pipeline produces semantically grounded editing instructions, achieving the highest NIMA and CLIP-IQA on both datasets. For content fidelity, PosterO and hpcvtg use frames from our Stage 1 retrieval, so they match our fidelity scores. Notably, KuaiHL and ShowMe exhibit worse fidelity, indirectly confirming the effectiveness of our retrieval model. For style personalization, most baselines score 0.00 as they produce identical thumbnails across users. Our method achieves IUD of 0.15 by leveraging the personalized intensity predictor. It assigns user-specific editing strengths based on style profiles, producing visually distinct thumbnails from same base frame. 

\vspace{-1em}
\subsection{Effectiveness of Components}
\label{sec:ablation}
To validate each component's contribution, we conduct ablation studies on EEGSVRec. Results are presented in Table \ref{tab:ablation_eegsvrec}. Removing any single module leads to consistent degradation. 

\noindent Removing \textbf{Personalized Highlight Retrieval (PHR)} replaces our retrieval model with random frame selection. This causes the largest fidelity drop. Randomly selected frames are poorly aligned with user-highlighted content. This confirms that Stage 1 provides the essential content anchor for downstream generation.

\noindent Removing \textbf{Personalized Thumbnail Generation (PTG)} uses the retrieved frame directly as the thumbnail. This yields the largest thumbnail quality drop. This demonstrates that generation is indispensable for both visual enhancement and inter-user diversity.

\noindent Removing \textbf{Personalized Intensity Prediction (PIP)} applies a uniform editing intensity for all users. This removal leads both thumbnail quality and style personalization decreases. This confirms that per-user intensity calibration provides orthogonal style-level personalization beyond content selection.

\begin{figure}[htbp]
    \centering
    \includegraphics[width=1.0\linewidth]{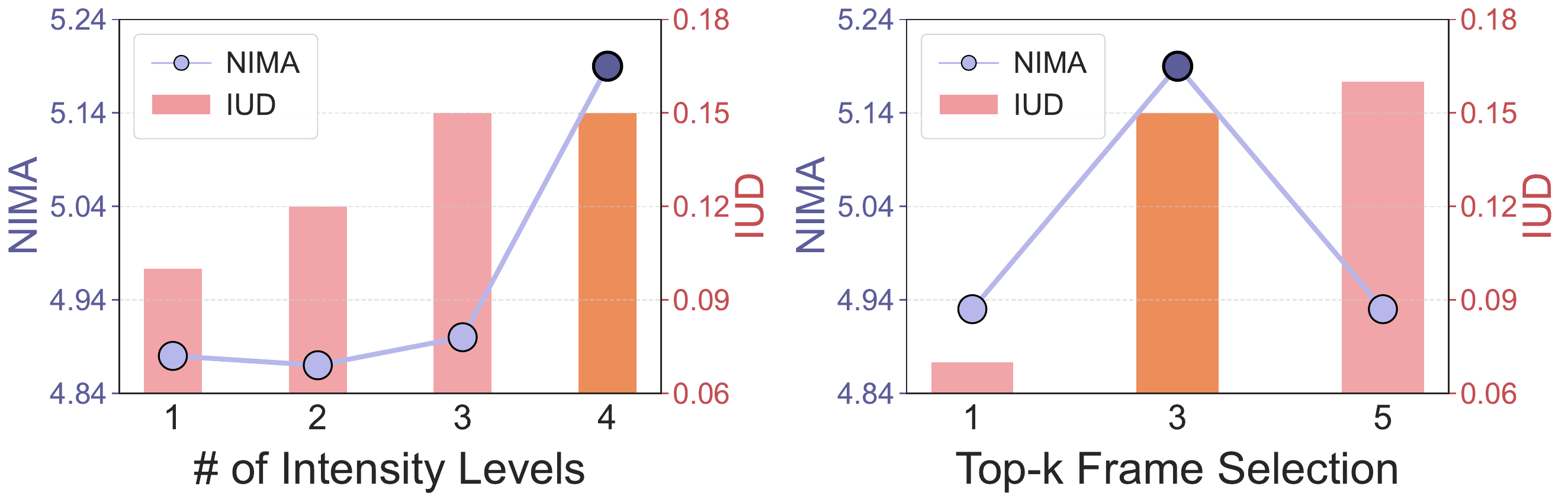}
    
    \setlength{\abovecaptionskip}{0.1cm}
\caption{Effect of the hyper-parameters on thumbnail quality~(NIMA) and style personalization~(IUD) on EEGSVRec.}
    \label{fig:placeholder}
\end{figure}

\subsection{Hyper-parameter Study}
We investigate two key hyperparameters in Stage~2 on EEGSVRec. We report NIMA and IUD as representative metrics. Content fidelity remains constant since these parameters only affect generation, leaving the retrieved base frame unchanged.

\noindent \textbf{Number of Intensity Levels.} Both NIMA and IUD increase as more levels are introduced. With only one level, all users receive identical editing strength. This yields the lowest style personalization. Each additional level provides finer style granularity. The intensity predictor can better match individual preferences from a richer candidate set. We use all four levels as default setting.

\noindent \textbf{Top-K Frame Selection.} When $k=1$, no supporting frames are available for cue injection, limiting both quality and diversity. At $k=5$, although IUD continues to increase, NIMA drops notably. This is likely because incorporating too many heterogeneous frames introduces conflicting visual elements during fusion, degrading thumbnail coherence. In contrast, $k=3$ achieves the best balance: it provides sufficient visual cues for enrichment while preserving coherence. We thus select $k=3$ as the default setting.
\section{User-Centric Evaluation}


Offline metrics can assess visual quality, but they do not answer a key question: \textit{do personalized thumbnails actually change user behavior?}
Since thumbnails primarily function as click triggers, their effectiveness should be evaluated through user interaction.
Thus, we design a user study that directly measures how different thumbnail generation strategies influence user click choices, with a focus on whether aligning thumbnails with individual preferences improves click preference.



\subsection{User Study Protocol}

\begin{figure}[htbp]
    \centering \setlength{\abovecaptionskip}{0.1cm}\includegraphics[width=8.5cm]{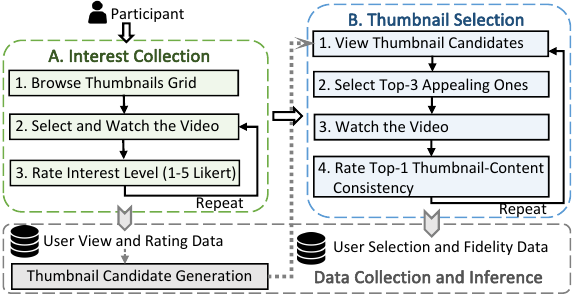}
    \caption{User study procedure with interest collection and thumbnail selection.}
    \label{fig:Userstudy_procedure} 
\end{figure}

\begin{figure*}[htbp]
    \centering
    \setlength{\abovecaptionskip}{0.cm}
    \includegraphics[width=17.8cm]{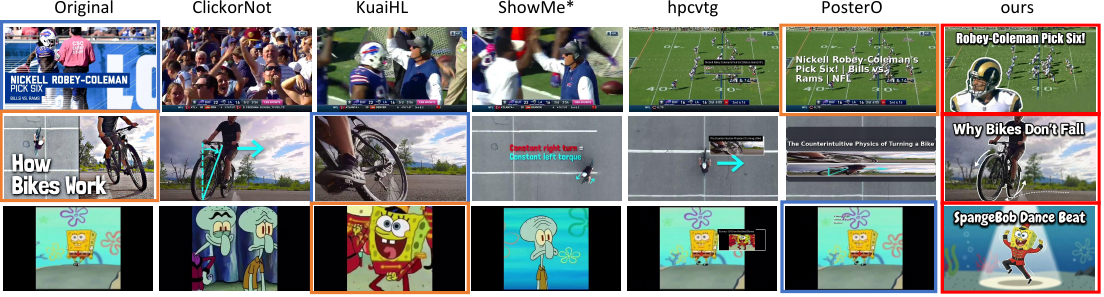}
    \caption{Qualitative comparison of thumbnails generated by different methods. Colored boxes indicate user click preference: \textcolor{red}{red} = 1st choice, \textcolor{orange}{orange} = 2nd choice, \textcolor{blue}{blue} = 3rd choice.}
    \label{fig:case_baseline} 
    \vspace{-1em}
\end{figure*}

Personalized thumbnail generation relies on user preference signals. To simulate a realistic usage scenario, we develop a video browsing interface and design a two-module user study~(Figure~\ref{fig:Userstudy_procedure}). This design enables direct measurement of click-through preference across generation strategies and injection intensity, and the two modules are introduced below: 

\noindent \textbf{Module A: Preference Collection.}
All participants were exposed to the same pool of \textbf{100 candidate videos}, each displayed with its original platform thumbnail to avoid generation bias. 
They browsed the video pool and selected \textbf{at least 20 videos} of personal interest. For each selected video, they watched it and rated their interest level on a 5-point scale. These ratings serve as the user preference signal for personalized retrieval and thumbnail generation in Module B.

\noindent \textbf{Module B: Thumbnail Evaluation.}
Using Module A ratings. \textbf{20 videos} were recommended per participant. For each video, participants were presented with thumbnails generated by our method at multiple injection-intensity levels alongside baseline thumbnails. 
Participants clicked the thumbnail they found most appealing, and then rated the consistency between the top-1 thumbnail and the video content after viewing the full video.

\subsection{Thumbnail-Driven Click Performance} 

\begin{table}[t]
\setlength{\abovecaptionskip}{0.1cm}
\renewcommand\arraystretch{1.1}
\caption{Comparison of different methods across Top-k choices in User Study. 'Orig.', 'CoN', and 'Post.' denote Original, ClickorNot, and PosterO, respectively.}
\label{tab:click_method}
\centering
\small 
\setlength{\tabcolsep}{3pt} 
\begin{tabular}{l | c c c c c c c}
\hline
\textbf{Choices} & 
\textbf{Orig.} & 
\textbf{CoN} & 
\textbf{ShowMe} & 
\textbf{KuaiHL} & 
\textbf{hpctvg} & 
\textbf{Post.} & 
\textbf{Ours} \\
\hline
Top1 & 7.60\%  & 7.60\%  & 8.30\%  & 7.40\%  & 1.20\%  & 4.50\%  & \textbf{63.30\%} \\
Top2 & 7.60\%  & 9.00\%  & 8.50\%  & 7.50\%  & 1.00\%  & 5.80\%  & \textbf{60.60\%} \\
Top3 & 8.00\%  & 10.20\% & 8.40\%  & 7.80\%  & 1.00\%  & 6.00\%  & \textbf{58.70\%} \\
\hline
\end{tabular}
\end{table}

\begin{table}[t] 
  \definecolor{lightpurple}{RGB}{248, 231, 242}
  \centering 
\setlength{\abovecaptionskip}{0.1cm}
  \caption{Quantitative comparison of different methods on user study dataset.}
  \label{tab:userstudy_metrics}
  \setlength{\aboverulesep}{0pt}
  \setlength{\belowrulesep}{0pt}
\small 
\setlength{\tabcolsep}{2.5pt} 
\begin{tabular}{l ccccc}
  \toprule
  \textbf{Method} & \textbf{NIMA} $\uparrow$ & \textbf{CLIP-IQA} $\uparrow$ & \textbf{DINO-SIM} $\uparrow$ & \textbf{LPIPS} $\downarrow$ & \textbf{IUD} $\uparrow$ \\
  \midrule
  Original   & 4.76 & \underline{0.74} & 0.46 & 0.57 & 0.00 \\
  ClickorNot~\cite{song2016click} & 4.75 & 0.73 & 0.44 & 0.56 & 0.00 \\
  KuaiHL~\cite{deng2024multimodal}     & 4.66 & 0.72 & \underline{0.45} & \underline{0.56} & 0.00 \\
  ShowMe~\cite{bhattacharya2022show}     & 4.62 & 0.69 & 0.42 & 0.57 & 0.01 \\
  PosterO~\cite{hsu2025postero}    & 4.77 & 0.69 & \textbf{0.60} & \textbf{0.41} & \underline{0.11} \\
  hpcvtg~\cite{yang2023toward}     & \underline{4.94} & 0.68 & \textbf{0.60} & \textbf{0.41} & 0.10 \\
  \midrule
  \rowcolor{lightpurple}
  \textbf{Ours} & \textbf{5.20} & \textbf{0.75} & \textbf{0.60} & \textbf{0.41} & \textbf{0.21} \\
  \bottomrule
\end{tabular}
\end{table}

\begin{figure}[htbp]
    \centering
    \setlength{\abovecaptionskip}{0.cm}
    \includegraphics[width=8.7cm]{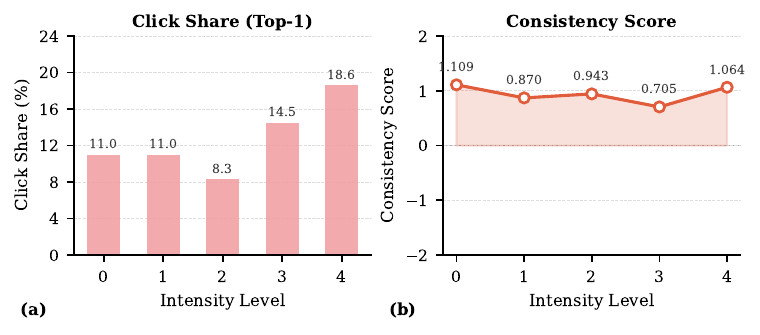}
    \caption{(a)~Click share and (b)~human fidelity rating of our method by editing intensity level under Top-$k$ choices.}
    \label{fig:click_consistency} 
\end{figure}

We recruited \textbf{30 participants} (age 18--35, diverse video consumption habits), all of whom completed both modules in full. Each participant was recommended personalized thumbnails of 20 videos. 


Table~\ref{tab:click_method} shows the proportion of top-$k$ selections for each method. Our approach achieves a dominant \textbf{63.30\%} Top-1 click ratio, far exceeding all baselines (the best baseline reaches 8.30\%). Moreover, each editing intensity in our method consistently achieves a higher click ratio than all baselines.
This gap suggests that thumbnail effectiveness depends less on generic visual quality and more on \textit{alignment with user preference}. Methods that produce a single universal thumbnail fail to capture this variability, while our personalized generation consistently ranks higher across Top-1/2/3.

To verify that offline metric advantages transfer to the user study, Table \ref{tab:userstudy_metrics} reports quantitative results on the same 30-user pool. Our method achieves the highest thumbnail quality while matching the best fidelity scores. Notably, our IUD nearly doubles the second-best baseline, indicating that our thumbnails are substantially more differentiated across users. These results directly align with the dominant click share reported in Table \ref{tab:click_method}.


To analyze the effect of editing intensity, Figure~\ref{fig:click_consistency} reports click ratio and human-rated consistency scores across intensity levels 0--4~(from no editing to extra-high). Click ratio reveals two distinct user preferences: some users favor lower intensities~(0–1) with more realistic thumbnails, while others prefer higher intensities~(3–4) with stronger visual enhancement, reflecting a divergence in style preference. 
Meanwhile, user-evaluated consistency scores remain stable across all levels, with ratings consistently distributed within the [-2, 2] range. This indicates that even at higher intensities, thumbnails preserve alignment with the source content, suggesting that stronger editing can enhance visual appeal without compromising content consistency.


\begin{figure}[htbp]
    \centering
    \setlength{\abovecaptionskip}{0.cm}
    \includegraphics[width=8.7cm]{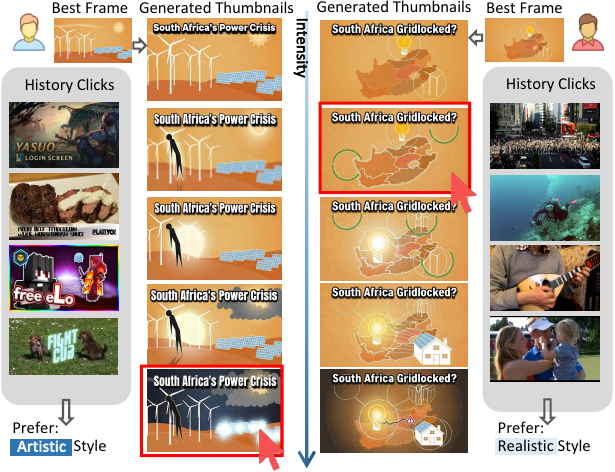}
    \caption{Effect of editing intensity on personalized thumbnail of the same video for two users with different click histories (artistic vs.\ realistic). Red boxes mark the model-predicted intensity levels, which are also selected by users.}
    \label{fig:case_persona} 
\end{figure}



\subsection{Thumbnail Generation Case Analysis}

To illustrate the practical benefits of personalization, Figure~\ref{fig:case_baseline} compares thumbnails generated by different methods. Our method consistently receives the highest-ranked selections, while baseline thumbnails are distributed across lower preferences or left unchosen. 
This pattern reflects the advantage of grounding generation in personalized visual cues rather than optimizing a generic objective, and is consistent with the quantitative results in Table~\ref{tab:click_method}.

Figure~\ref{fig:case_persona} highlights the role of the editing intensity prediction. Users with a preference for stylized, artistic content are assigned higher intensity, producing visually enhanced thumbnails, while users favoring realistic content receive lower intensity, maintaining closer to the source frame. These cases show that the predictor successfully maps diverse style preferences to appropriate editing intensity, allowing the same generation pipeline to serve different user tastes while preserving content consistency.




\section{Conclusions and Future Work}

We present the first study on personalized video thumbnail generation, introducing a two-stage framework that combines highlight retrieval with VLM-guided diffusion editing under personalized intensity control. Our method achieves superior retrieval accuracy, generation performance in the quantitative experiments, and click share in the user study, demonstrating the effectiveness of integrating user preference signals into visual generation.
Future work includes exploring end-to-end optimization to jointly learn retrieval and generation, and deploying the approach in real-world platforms to measure actual engagement gains beyond user study and offline experiments.

\clearpage
\bibliographystyle{ACM-Reference-Format}
\bibliography{reference}


\clearpage

\appendix

\section{Datasets}

\subsection{Dataset Description}

We conduct experiments on two public datasets that provide user-specific shot-level preference annotations:

\textbf{EEGSVRec}~\cite{zhang2024eeg} is a short-video recommendation dataset that records users' EEG signals while watching videos, from which per-user interest levels are derived using a standard neurophysiological engagement index~\cite{pope1995biocybernetic}, providing implicit yet fine-grained interest labels.

\textbf{PHD$^2$}~\cite{garcia2018phd} is a personalized highlight detection dataset constructed from users on gifs.com who create GIFs by selecting segments from full YouTube videos. The timestamps of these user-generated GIFs are treated as explicit annotations of user-specific highlights, providing large-scale supervision of individual viewing preferences.

\begin{table}[htbp]
\small
\renewcommand\arraystretch{1.2}
\setlength{\abovecaptionskip}{0.cm}
\setlength{\belowcaptionskip}{0.cm}
\caption{Datasets statistics.}
\centering
\begin{tabular}{p{1.1cm}|p{0.8cm}<{\centering}p{0.8cm}<{\centering}p{0.8cm}<{\centering}p{1.0cm}<{\centering}p{1.4cm}<{\centering}@{}}
\hline
\textbf{Datasets} & \textbf{\#Users} &
\textbf{\#Items} & \textbf{\#Views} & \textbf{\#Highlights} & \textbf{\#avg Shots}\\
\hline
EEGSVRec & 30 & 2,636 & 3,657 & 2.23 & 26.47 \\
PHD$^2$ & 12,973 & 63,990 & 112,014 & 2.96 & 58.36\\
\hline
\end{tabular}
\label{tab:dataset}
\end{table}

For EEGSVRec, raw EEG signals are preprocessed as follows: per-second differential entropy is computed for the theta ($4$--$8$\,Hz), alpha ($8$--$13$\,Hz), and beta ($13$--$30$\,Hz) bands, and a per-second engagement index $e = \beta / (\theta + \alpha)$ is computed following~\cite{pope1995biocybernetic,ahmed2025eeg}. Scores are then z-score normalized within each user, and the highlight sub-segment is defined as the continuous interval exceeding the per-video 60th-percentile engagement for at least 2 seconds; among all qualifying intervals, the one with the highest mean engagement is selected as the ground-truth highlight. We partition both datasets using an 8:1:1 user-based split for training, validation, and testing. The dataset statistics are shown in Table~\ref{tab:dataset}. 

\subsection{Datapreprocess Details}

Personalized highlight retrieval requires video representations that are both semantically meaningful and temporally structured. User interest naturally aligns with scenes and moments rather than fixed-frame intervals. We therefore represent each video $v \in \mathcal{V}$ as an ordered sequence of shots ${s_1, \ldots, s_N}$, detected by PySceneDetect~\cite{pyscenedetect}, where each shot forms a temporally coherent and semantically consistent unit. 

From each shot $s_i$, a single representative key frame $f_i$ is selected for both interest scoring and thumbnail generation. Candidate frames are first identified via CLIP-based adaptive clustering~\cite{dhiman2023keyframe,tan2024keyframe}, then ranked by a composite score considering semantic diversity, visual quality, aesthetics, face ratio, contrast, and sharpness. Both the selected key frames and the video caption (title and tags) are encoded with CLIP~\cite{radford2021learning}.

\section{Baseline Details}
\label{app:baselines}

\noindent \textbf{(i) Key Frame Selection Methods.} These types of baselines are derived from the video highlight detection task:
\textbf{HighlightMe}~\cite{bhattacharya2021highlightme} is a non-personalized highlight detector that identifies highlights without user-specific annotations.
\textbf{PAC-Net}~\cite{garcia2018phd} conditions a content-based detector on user-annotated GIF preferences.
\textbf{PR-Net}~\cite{chen2021pr} constructs a preference reasoning graph to guide personalized highlight selection.
\textbf{ShowMe}~\cite{bhattacharya2022show} employs multi-head attention over content features and user history for personalized highlight detection.
\textbf{KuaiHL}~\cite{deng2024multimodal} uses a multimodal Transformer to predict frame-level engagement in short-video scenarios.
\textbf{ClickorNot}~\cite{song2016click} is specifically designed for thumbnail selection by scoring frames with learned quality predictors, but produces a single generic thumbnail shared across all users.

\noindent \textbf{(ii) Layout Generation Methods.} \textbf{HPCVTG}~\cite{yang2023toward} generates thumbnails following aesthetic and perception principles refined with human feedback.
\textbf{PosterO}~\cite{hsu2025postero} leverages LLM-based in-context learning to produce content-aware poster layouts.
In our experiments, both methods are supplied with frames retrieved by our Stage~1 model for fair comparison.

\section{Evaluation Metric Details}
\label{app:metrics}

\subsection{Highlight Retrieval Metrics}
We evaluate per-frame ranking quality from three complementary angles. \textbf{mAP} averages precision over all positive ranks per video and then over videos, capturing overall ranking quality independent of any cutoff. \textbf{Recall@$k$} ($k \in \{1,3,5,10\}$) measures the fraction of ground-truth highlight frames retrieved in the top-$k$, with $k{=}1$ reflecting strict thumbnail-selection accuracy and larger $k$ tolerating near-misses. \textbf{nMSD} (normalized Mean Squared Displacement) computes $\frac{1}{|R|}\sum_{i \in R}(\text{rank}(i) - \text{ideal}(i))^2$ normalized by the worst-case displacement, penalizing relevant frames that drift far from their ideal positions. nMSD complements Recall@$k$ by remaining sensitive to ranking errors that fall just outside the top-$k$ window.

\subsection{Thumbnail Generation Metrics}
A personalized thumbnail must satisfy three orthogonal criteria: it should be visually appealing, faithful to the source video, and distinct across users. Optimizing any single axis is insufficient. A thumbnail can be aesthetic but generic, or personalized but hallucinated, so we report all three.

\noindent\textit{Thumbnail Quality.} \textbf{NIMA}~($\uparrow$)~\cite{talebi2018nima} predicts mean opinion scores from a CNN trained on the AVA aesthetics dataset, capturing low-level perceptual appeal. \textbf{CLIP-IQA}~($\uparrow$)~\cite{CLIP-IQA} complements NIMA by scoring semantic-level quality through contrastive prompts, which is sensitive to compositional coherence rather than only sharpness or exposure. Both metrics are computed on text-stripped thumbnails to prevent caption overlays from dominating the score.

\noindent\textit{Content Fidelity.} For each (user, video) pair, we compute the metric between the method's selected key or base frame and \emph{each} ground-truth highlight frame annotated for that user, then average. This protocol rewards methods whose output remains close to any of the user's highlighted moments, not only a single anchor. \textbf{LPIPS}~($\downarrow$)~\cite{LPIPS} measures perceptual distance via deep feature differences, while \textbf{DINO-Sim}~($\uparrow$) computes cosine similarity over DINOv2~\cite{DINO} embeddings, which capture object-level structure that LPIPS's low-level features miss. The two are reported jointly because perceptual closeness and structural correspondence can diverge under heavy editing.

\noindent\textit{Style Personalization.} Existing metrics measure absolute quality but cannot detect whether different users actually receive different thumbnails. We therefore propose \textbf{IUD} (Inter-User Dissimilarity, $\uparrow$): for a video $v$ viewed by user set $U_v$ with $|U_v| \geq 2$,
\begin{equation}
\text{IUD}(v) = \frac{2}{|U_v|(|U_v|-1)} \sum_{i<j}\bigl(1 - \cos(z_i, z_j)\bigr),
\end{equation}
where $z_i$ is the CLIP ViT-L/14 embedding of user $u_i$'s generated thumbnail $\hat{T}(u_i, v)$. We average IUD over all qualifying videos. CLIP embeddings are used because they jointly encode semantic and stylistic variation, whereas pixel-level distances would conflate genuine personalization with sampling noise.

\section{Implementation Details for Personalized Thumbnail Generation}
\label{sec:pipeline-walkthrough}

Stage 2 takes the top-$K$ frames retrieved by Stage 1, the video metadata $m$, and the user click history $\mathcal{H}^T_u$, and produces a personalized thumbnail $\hat{T}(u, v)$. Algorithm~\ref{alg:pipeline} summarizes the full procedure, and the three core components are detailed in Sections~\ref{sec:cue_extraction}--\ref{sec:verify-retry}: structured visual cue extraction (\S\ref{sec:cue_extraction}), editing intensity prediction (\S\ref{sec:intensity-prediction}), and the verify-and-retry loop (\S\ref{sec:verify-retry}). 

Three design choices are worth noting. First, frame cleaning precedes all downstream steps because residual text and watermarks would otherwise be inherited by both the cue extractor as spurious content and the diffusion editor as visual artifacts. Second, ``what to edit'' and ``how much to edit'' are decoupled into two independent modules: cue extraction derives content-level edits from the retrieved frames and video metadata, while intensity prediction derives style-level edits from the user's click history. This separation allows the same extracted plan to serve users with different stylistic preferences by simply switching the intensity filter. Third, the verify-retry loop resamples only the diffusion noise across attempts while holding the base frame, supporting frames, and filtered plan $\pi_u$ fixed, which isolates stochastic generation failures from plan-level errors. 

Runtime is dominated by the VLM and diffusion calls: Qwen cue extraction takes roughly 60 seconds, each diffusion editing attempt takes around 30 seconds, and each judge call takes about 60 seconds, consistent with the $\sim$210 second per-video aggregate reported in Section~\ref{implementation}.

\begin{algorithm}[t]
\renewcommand{\algorithmicrequire}{\textbf{Input:}}
\renewcommand{\algorithmicensure}{\textbf{Output:}}
\caption{Stage 2: Personalized Thumbnail Generation Pipeline}
\label{alg:pipeline}
\begin{algorithmic}[1]
\Require User $u$, video $v$, top-$K$ retrieved frames $\{s_1, \dots, s_K\}$ from Stage 1,
         video metadata $m$, user click history $\mathcal{H}_u^T$
\Ensure  Personalized thumbnail $\hat{T}(u, v)$

\Statex \hspace{-1.2em}\textcolor{gray}{\textit{// Frame Preprocessing}}
\For{$k = 1$ \textbf{to} $K$}
    \State $s_k' \gets \Call{DiffClean}{s_k}$
        \Comment{remove overlaid text/watermarks}
    \State $\tilde{s}_k \gets \Call{VLM-Select}{s_k'}$
        \Comment{select cleanest variant~(\S B.3)}
\EndFor

\Statex \hspace{-1.2em}\textcolor{gray}{\textit{// Structured Visual Cue Extraction~(\S B.1)}}
\State $\mathcal{P} \gets \Call{ExtractCues}{\tilde{s}_1, \dots, \tilde{s}_K,\; m}$
        \Comment{content summary, fusion decision, operations}
\State $f_{\text{base}} \gets \mathcal{P}.\textsc{base}$;\quad
       $\mathcal{S}_{\text{support}} \gets \mathcal{P}.\textsc{support}$

\Statex \hspace{-1.2em}\textcolor{gray}{\textit{// Editing Intensity Prediction~(\S B.2)}}
\State $\mathbf{q}^{(u)} \gets \Call{StyleProfile}{\mathcal{H}_u^T}$
        \Comment{4-D CLIP style vector}
\State $\lambda_u \gets \arg\min_{l \in \mathcal{L}}
       \lVert \mathbf{q}^{(u)} - \mathbf{p}^{(l)} \rVert_2$
        \Comment{nearest intensity prototype}
\State $\pi_u \gets \Call{FilterOps}{\mathcal{P}.\textsc{ops},\, \lambda_u}$
        \Comment{retain permitted operations}

\Statex \hspace{-1.2em}\textcolor{gray}{\textit{// Controllable Diffusion Editing with Verify-Retry~(\S B.3)}}
\State $\hat{C} \gets \varnothing$;\quad $\sigma^{\star} \gets -\infty$
\For{$r = 1$ \textbf{to} $N$}
    \State $\hat{T}_r \gets \Call{DiffEdit}{f_{\text{base}},\, \mathcal{S}_{\text{support}},\, \pi_u}$
        \Comment{resampled noise per attempt}
    \State $\sigma_r \gets \Call{VLM-Judge}{\hat{T}_r,\, \pi_u}$
    \If{$\sigma_r > \sigma^{\star}$}
        \State $\hat{C} \gets \hat{T}_r$;\quad $\sigma^{\star} \gets \sigma_r$
    \EndIf
    \If{$\sigma_r \geq \tau$}
        \State \textbf{break}
    \EndIf
\EndFor

\Statex \hspace{-1.2em}\textcolor{gray}{\textit{// Text Overlay}}
\State $\hat{T}' \gets \Call{DiffText}{\hat{C},\, \mathcal{P}.\textsc{caption}}$
\State $\hat{T} \gets \Call{VLM-Select}{\hat{T}'}$
\State \Return $\hat{T}$
\end{algorithmic}
\end{algorithm}



\subsection{Structured Visual Cue Extraction}
\label{sec:cue_extraction}

\noindent\textbf{Inputs.} Following Algorithm~\ref{alg:pipeline}, the cue extractor receives the cleaned top-3 key frames retrieved by Stage~1 together with the video's topic tag.

\noindent\textbf{Prompt design.} We prompt Qwen3.5-27B~\cite{qwen_qwen3_5_27b_hf} with a task framing that treats the top-1 frame as the primary semantic anchor and the remaining frames as optional supporting references. The prompt instructs the model to make an explicit fusion decision: if supporting frames carry complementary topic-relevant cues, the model specifies selective fusion; otherwise, it must strengthen topic-level visual communication through generated elements rather than defaulting to surface beautification. The prompt also defines the four-level intensity hierarchy (low, medium, high, extra-high) as a strict cumulative superset, so that the downstream intensity predictor in Section~\ref{sec:intensity-prediction} can select among pre-compiled plans without re-querying the VLM. The full prompt template is released with our code.

\noindent\textbf{Output schema.} The structured editing plan is organized into four functional blocks; the full schema is documented in the code release. \emph{Semantic analysis} (\texttt{anchor\_summary}, \texttt{semantic\_similarity\_assessment}, \texttt{top1\_topic\_coverage}) captures the VLM's understanding of the top-1 frame as the primary anchor and assesses whether supporting frames carry complementary cues. \emph{Frame routing} (\texttt{fusion\_decision}, \texttt{editor\_input\_plan}) determines which frame becomes the base and which supporting frames contribute visual cues. \emph{Edit operations} are decomposed into five categories: topic visualization, composition, subject emphasis, background control, and a dedicated \texttt{supporting\_fusion\_edits} field for cross-frame fusion. Finally, \emph{execution prompts} (\texttt{stage1\_execution\_prompts}) compile these operations into four ready-to-use instructions at increasing intensities, which form the candidate set selected from in Section~\ref{sec:intensity-prediction}.

\noindent\textbf{Worked example.} Figure~\ref{fig:cue_example} shows a representative output for a horoscope vlog whose top-3 frames are visually similar shots of a presenter waving at the camera. The model judges semantic similarity as high and top-1 topic coverage as weak, so it routes to \texttt{fusion\_decision.mode = none} and instead populates \texttt{topic\_visualization\_edits} with imagined elements (zodiac constellation icons, starry background texture, mystical glow) that ground the generation in the video's topic rather than its raw frames. The four execution prompts then layer these edits cumulatively: \emph{low} adds only subject emphasis and background blur; \emph{medium} introduces composition shifts; \emph{high} injects the imagined topic elements; \emph{extra-high} additionally transforms the background atmosphere. This example illustrates how the cue extractor produces semantically meaningful edits even when no supporting frame fusion is possible, grounding the downstream diffusion edits in user-preferred content rather than arbitrary design choices.

\begin{figure}[h]
\centering
\begin{lstlisting}[basicstyle=\scriptsize\ttfamily, breaklines=true, frame=single]
"fusion_decision": {
  "mode": "none",
  "reason": "The three frames are semantically identical
    in subject and setting. No additional cues from top2
    or top3 are needed; new thematic elements must be
    generated to address the 'horoscope' topic.",
  "expected_benefit": "Maintaining the single strongest
    frame ensures the subject remains the focal point,
    allowing generated thematic overlays without clutter."
},
"topic_visualization_edits": [
  {"rank": 1,
   "element": "Floating zodiac constellation icons",
   "placement": "upper area and right side",
   "source": "imagined"},
  {"rank": 2,
   "element": "Subtle starry galaxy background texture",
   "placement": "overlaid on background",
   "source": "imagined"},
  {"rank": 3,
   "element": "Soft glowing mystical aura",
   "placement": "around subject head and shoulders",
   "source": "imagined"}
],
"stage1_execution_prompts": {
  "low":    "Apply a subtle soft white glow around the
             subject. Sharpen the face. Blur the background
             moderately.",
  "medium": "... + Enlarge the subject to dominate the
             center. Shift the subject slightly left to
             open space on the right side.",
  "high":   "... + Add floating zodiac constellation icons
             in the upper area and right side. Tint the
             background with a deep blue hue.",
  "extra_high": "... + Overlay a starry galaxy texture.
             Tint background deep purple. Add a mystical
             glowing aura around the subject's head."
}
\end{lstlisting}
\caption{Worked example of a structured editing plan for a horoscope vlog. The fusion decision rejects supporting frames in favor of imagined topic elements, which are then layered cumulatively across the four execution prompts. Ellipses (\texttt{...}) indicate inheritance from the previous intensity level.}
\label{fig:cue_example}
\end{figure}

\subsection{Editing Intensity Prediction}
\label{sec:intensity-prediction}

This section provides implementation details that complement Section~\ref{Intensity Prediction}, covering the offline calibration phase that builds a prototype for each intensity level. The three paragraphs below describe the contrastive prompt design, the reference thumbnail curation procedure, and the prototype design rationale.

\noindent\textbf{Style axis prompts.} Each of the four style axes is described by a contrastive low/high prompt pair, listed in Table~\ref{tab:style-prompts}. We chose these four axes specifically because they cover the orthogonal dimensions that distinguish realistic from artistic thumbnails: focal salience (pop-out), framing tightness (dominance), informational layering (cue density), and atmospheric treatment (stylization). We encode all prompts and thumbnails with a frozen CLIP ViT-L/14, and compute each axis score as the difference between cosine similarities to the high and low prompts.

\noindent\textbf{Reference thumbnail curation.} For each intensity level $l \in \mathcal{L}$, we generate 50 reference thumbnails by running our own editing pipeline at the corresponding intensity, with source frames randomly sampled from the EEGSVRec~\cite{zhang2024eeg} and PHD$^2$~\cite{garcia2018phd} training sets. Generating references with the same editor that the predictor will later select intensities for ensures that the reference distribution matches the deployment distribution, avoiding calibration drift between offline prototypes and online generations. We found $K=50$ empirically sufficient for stable centroids.

\noindent\textbf{Prototype design.} Unlike a single-axis intensity scale, our four intensity levels are not scaled versions of one another. As defined in Section~\ref{Intensity Prediction}, each level targets a distinct editing signature: low operates only on surface treatments, medium adds layout adjustments, high additionally injects cross-frame fusion cues, and extra-high further introduces scene transformation. Because each level activates a different combination of edit types, its prototype vector occupies a distinct region of the 4D style space rather than sitting at a higher position along a single axis. This is why nearest-neighbor matching is the right inference rule: a user's style profile is matched to the signature pattern it most resembles, not to a position on a one-dimensional intensity scale.


\begin{table}[t]
\centering
\caption{Contrastive low/high prompt pairs for the four style axes used in CLIP-based scoring.}
\label{tab:style-prompts}
\small
\begin{tabular}{p{0.18\linewidth} p{0.36\linewidth} p{0.36\linewidth}}
\toprule
\textbf{Axis} & \textbf{Low prompt} & \textbf{High prompt} \\
\midrule
Subject pop-out &
a casual frame with no strong focal point &
a cover image with a single clear focal point and strong subject emphasis \\
\addlinespace
Subject dominance &
a thumbnail showing the subject from far away in a wider shot &
a thumbnail showing the subject in a tight crop or close-up that fills much of the frame \\
\addlinespace
Semantic cue density &
a very sparse cover with almost no supporting items &
a layered cover with several supporting items that explain the topic \\
\addlinespace
Scene stylization &
a thumbnail with ordinary lighting and a realistic background &
a thumbnail with strong mood lighting, vivid color contrast, and a visually designed background \\
\bottomrule
\end{tabular}
\end{table}

\subsection{Verify-and-Retry Loop}
\label{sec:verify-retry}

The verify-and-retry loop guards against stochastic diffusion failures, such as hallucinated objects or partial execution of the structured editing plan, by re-invoking the editor with a fresh noise sample when a fidelity check fails. 

\noindent\textbf{Judge prompt design.} The VLM judge (Qwen3.5-27B~\cite{qwen_qwen3_5_27b_hf}) takes the source frame, the edited thumbnail, and the intensity-filtered editing plan $\pi_u$ as input. It is instructed by its prompt to first identify 1--4 atomic edit requirements from $\pi_u$ and then evaluate each independently, plus two global checks for visual coherence and subject preservation. Anti-hallucination instructions in the prompt template require all judgments to be grounded in evidence visible in the edited image, preventing the judge from accepting generations based on assumed rather than observed content.


\noindent\textbf{Scoring schema.} For each generated thumbnail, the judge outputs a single discrete rating drawn from \{Poor, Fair, Good, Very Good, Perfect\}, which we map to integers $\{0, 1, 2, 3, 4\}$ to obtain the fidelity score $\sigma$. If $\sigma$ exceeds the acceptance threshold, the thumbnail is accepted; otherwise, the loop retries with a new noise sample. Subject preservation is the only hard-fail check, applied solely at the image-editing stage: if the judge marks subject preservation as failed, the overall rating is capped at Fair regardless of other checks, preventing the loop from accepting generations that lose the user's anchor content.

\noindent\textbf{Acceptance rule.} We set the acceptance threshold to $\tau = 3$ (Very Good), chosen empirically as the best balance between fidelity and accept rate. If the first attempt already satisfies $\sigma \geq \tau$, the loop terminates immediately without retrying. Otherwise, the diffusion editor is re-invoked with a new random noise sample while the base frame, supporting frames, and filtered plan $\pi_u$ are held fixed. The retry budget $N = 1$ is chosen as the best trade-off between effectiveness and efficiency, since each retry adds roughly 90 seconds (60s judge + 30s diffusion). If no attempt reaches $\tau$ within $N$ retries, the candidate with the highest $\sigma$ is returned.

\section{User Study Platform Workflow}

The user study was conducted via a web-based platform comprising two modules.
Participants were first assigned to either Module~A (preference collection) or Module~B (thumbnail evaluation), determined by pre-assigned credentials.

\begin{figure}[htbp]
    \centering
    \includegraphics[width=1.0\linewidth]{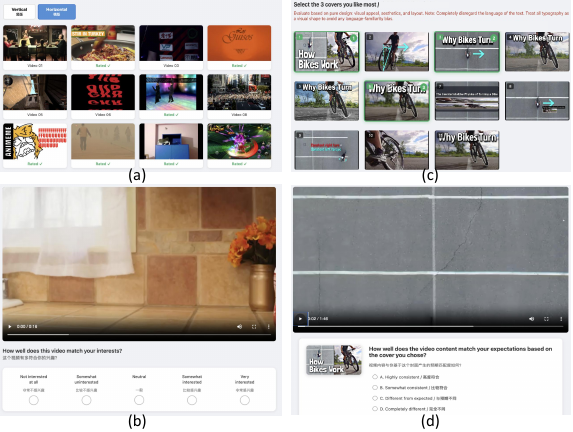}
    \setlength{\abovecaptionskip}{0.1cm}
\caption{User study platform interface. Module A: (a) grid view and (b) video player with a Likert scale; Module B: (c) thumbnail selection grid and (d) video player with a matching question.}
    \label{fig:platform}
\end{figure}

\subsection*{Module A: Video Preference Collection}

\begin{enumerate}[label=(\arabic*), leftmargin=*]
  \item \textbf{Login.} Participants log in with a pre-assigned user ID and password.
  \item \textbf{Language selection.} Participants select their primary language (Chinese or English). English speakers are additionally asked whether they can understand Chinese-language videos.
  \item \textbf{Instructions.} A task-instruction page is displayed for at least 10 seconds before the participant may proceed.
  \item \textbf{Video rating.} A grid of videos is presented, as shown in Figure~\ref{fig:platform}-(a). Participants click on any video to open the player. Seeking/scrubbing is disabled to encourage genuine viewing. After watching, participants rate the video on a 5-point Likert scale:
    \begin{itemize}
      \item 1 -- Not interested at all
      \item 2 -- Somewhat uninterested
      \item 3 -- Neutral
      \item 4 -- Somewhat interested
      \item 5 -- Very interested
    \end{itemize}
    Figure~\ref{fig:platform}-(b) is the interface of this process. The rating is saved and the participant returns to the grid to select the next video. The required minimum is 10 vertical-format and 10 horizontal-format videos for Chinese-proficient participants, or 20 horizontal-format videos for English-only participants.
\end{enumerate}

\subsection*{Module B: Thumbnail Evaluation}

\begin{enumerate}[label=(\arabic*), leftmargin=*]
  \item \textbf{Login \& instructions.} Same login flow; a separate instruction page describes the thumbnail-selection task.
  \item \textbf{Task list.} Each participant is pre-assigned 20 video cases. The task list shows completion status for each case.
  \item \textbf{Thumbnail selection.} For each case, approximately 11 candidate thumbnails (generated by different methods) are presented in shuffled order. Figure~\ref{fig:platform}-(c) is an example of the interface. The participant selects their top-3 preferred thumbnails in ranked order (1st, 2nd, 3rd choice) based solely on visual appeal, design quality, and layout---irrespective of language.
  \item \textbf{Video viewing \& match question.} After confirming the selection, the participant watches the corresponding video and answers one question, as shown in Figure~\ref{fig:platform}-(d):
    \begin{itemize}
      \item \textbf{Q} (single choice): \textit{``How well does the video content match your expectations based on the cover you chose?''} (A. Highly consistent / B. Somewhat consistent / C. Different from expected / D. Completely different)
    \end{itemize}
  \item \textbf{Repeat.} Steps 3--4 are repeated for all 20 assigned cases, after which the participant submits all responses.
\end{enumerate}
 

\end{document}